# Neutron Production & ZDC acceptance

ZDC, BRAN and Sci rate issues at Injection Energy

Sebastian White

*This program calculates the reduction in counting rate of the TAN detectors (ZDC,BRAN,Sci) at low beam energy due to the increased angular spread of the leading baryons. At $\sqrt{s} = 0.9$ TeV, $L = 10^{29}$ cm$^{-2}$ s$^{-1}$ and assuming that all detctors are sensitive to signals above 50 GeV the counting rates are ~100 Hz for single arm and ~4 Hz for coincidences.*

**Introduction:**

We calculate the effective cross section seen by the TAN detectors where $\sigma_{\mathrm{Eff}} = \left(\sigma_{2\text{-Arm}}{}^{\mathrm{ZDC}} \cdot \mathrm{Acceptance}\right)$ and find $\sigma_{\mathrm{Eff}}$=0.04, 0.35, 1.4 mb at $\sqrt{s}$=0.9, 2.0 and 4.0 TeV. At injection energy the signal is so small that uncorrelated backgrounds, mostly from $\gamma$'s, could be an issue. We also calculate the accidental background due to known physics.

This result also applies to the BRAN and SCI devices installed in the TAN by the LHC since they have the same acceptance and would also be operated as counters.

Since the calculated coincidence count rates are only a couple Hz, even once the LHC reaches L=10 $^{29}$, the only practical signal for beam tuning and luminosity monitoring is an "or" of the 2 arms, which we also calculate. Beam-gas backgrounds, potentially a problem for the "or" signal, are estimated.

This calculation is based almost exclusively on available data, including some from RHIC, since there are disagreements with available generators for pp.

## The detectors:

The three detectors occupy the same space in the TAN absorber. They are centered on the $0^o$ direction with respect to ATLAS interactions except at high luminosity, with crossing angle, when they are offset by 2 cms.

They all measure shower secondaries produced in the ZDC. BRAN and Sci are located after the first and second ZDC module $(1.2 \times \Lambda_{\mathrm{Int}})$ so they only see hadrons. ZDC sees mostly neutrons but also low energy $\gamma$'s which are expected to be uncorrelated. The ZDC has the best timing and energy resolution so it has no problem with low energy showers. Because of this it is expected to provide a very clean luminosity scaler, using coincidences, when $\sqrt{s} \geqslant 2$ TeV.

At $\sqrt{s}$ =14 TeV the BRAN has signal to noise better than 1:1 for a typical event so it can count single events and make coincidences. This wouldn't be possible at injection energy.

During the brief run of LHCf, when the first ZDC module has to be out, it would be impossible to say anything so definite. Part of the time LHCf is out of the beam so BRAN is insensitive and part of the time LHCf places an absorber with irregular shape directly in front of both BRAN and ZDC. It would be difficult to use these data except, possibly, for beam tuning.

## $p_T$ Acceptance:

Differential cross sections for inclusive (single) neutron production were measured at ISR[1], FNAL[2], SPS[3] and HERA[4].

PHENIX has preliminary measurements of both inclusive and coincident (2 neutron) cross sections at $\sqrt{s}$ =200 and 500 GeV.

About the measurement which matters most for calculating ZDC rates-$p_T$-there are discrepancies. NA49 doesn't give a $p_T$dependence but ISR and FNAL do. Over the range $0.3 < x_F < 0.7$ , ISR finds a constant slope for the invariant cross section:$E \frac{d^3 \sigma}{dp^3} \propto e^{-4.8\, p_T}$. It is clear from their plots that ISR gets a steeper $p_T$ slope for $x_F$>0.7 although they don't fit it. In this range we might expect a component of One Pion Exchange (OPE) production, which has a steeper $p_T$dependence, so we will approximate their results by combining the 2 forms in the high $x_F$region.

Note that PYTHIA doesn't include this. Only RAPGAP does. Hannes Jung, author of RAPGAP, is looking into its applicability to this problem.

ZEUS measured inclusive neutron production in e+p->n+X. Their results agree to better than a factor of ~2 with NA49 and ISR when plotted as $\frac{1}{\sigma^{\mathrm{inel}}} \frac{d\sigma}{dx_F}$.

Their results are fit with the form $e^{-b p_T^2}$ or $e^{bt}$ . Their b values suggest an increase with $x_F$ but this is not statistically very significant. Typically b is ~8 $(\mathrm{GeV}/c)^{-2}$ for most of the range.

Below we give the 3 $p_T$ forms and their integrals. We also show the final form we use to include the steeper $p_T$ slope at large $x_F$ indicated by both ISR and HERA data.

```
Needs["PlotLegends`"]

mπ = ParticleData["PiPlus", "Mass"] / 1000;
```

$$\text{OPE[pt_]} := \frac{3.842\times 10^{-4}}{\left(\text{pt}^2+\text{m}_\pi{}^2\right)^2}$$

$$\text{OPEInt[y_]} = \int_0^y \text{pt}\,\frac{.0393}{\left(\text{pt}^2+\text{m}_\pi{}^2\right)^2}\,d\text{pt};$$

$$\text{ISR[pt_]} := e^{-4.8\,\text{pt}}$$

$$\text{ISRInt[y_]} = 23.1\int_0^y \text{pt ISR[pt]}\,d\text{pt};$$

$$\text{HERA[pt_]} := e^{-8*\text{pt}^2}$$

$$\text{HERAInt[y_]} = 16\int_0^y \text{pt HERA[pt]}\,d\text{pt};$$

$$\text{Final[y_, x_]} = \text{If}\left[x > 0.75,\ \frac{\text{OPEInt[y]} + \text{ISRInt[y]}}{2},\ \text{ISRInt[y]}\right];$$

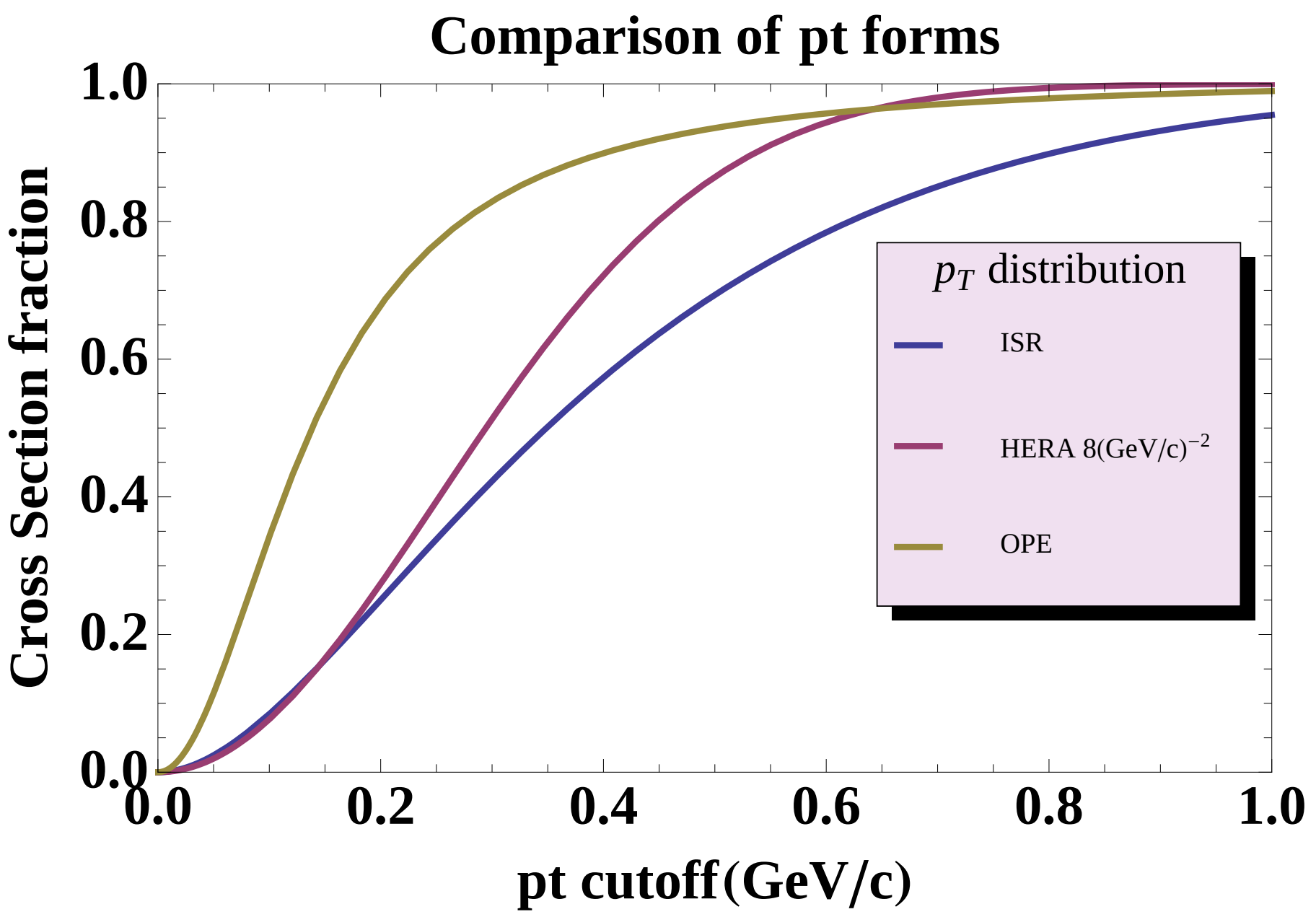

## Detector Acceptance:

Only a small part of the $p_T$ distribution is seen by the ZDC since pt = $x_F \times p_{\text{beam}} \times \theta$ and $\theta$<290 $\mu$rad assuming a fiducial cut of 4 cm on the neutron impact radius. The actual ATLAS acceptance is a little more complicated since the machine aperture is smaller than the ZDC size but 4cm is a close approximation.  For a 100 GeV neutron the $p_T$ cutoff is only 0.03 GeV. The geometry is illustrated below.

```
cm = 1/100;
fidcut = 4 cm/140;
Accept[pl_, x_] := Final[pl * fidcut, x]
```

neutron impact and $p_T$ cutoff

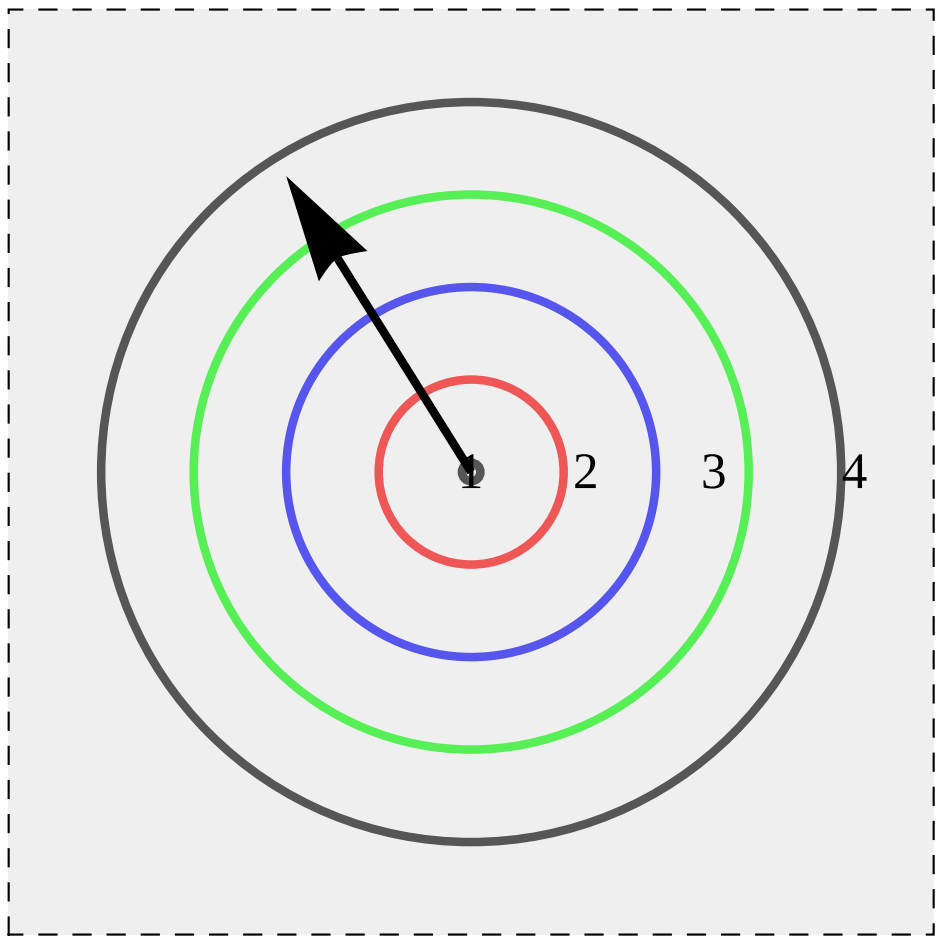

ZDC area(10×10 cms)and $p_T$ bins

## Yield:

The corrected inclusive distribution is the (average acceptance)×(fraction of $x_F$ range) multiplied by $< \frac{1}{\sigma_{\text{inel}}} \frac{d\sigma}{dx_{x_F}} >$.

We use $\frac{d\sigma}{dx_F} = \frac{(2\pi)}{x_F} \int E \frac{d^3\sigma}{dp^3} p_t \, dp_t$ to obtain it from the invariant cross section and make the approximation that the $x_F$ density is constant at ~ 0.4, roughly in agreement with the NA49 result (below). Note that ISR, HERA and NA49 don't agree that well on the average density and there is a well known factor of 2 spread.

For $\sigma_{inel}$ we use $\sigma_{total} - \sigma_{elastic} - \sigma_{SD}$ when calculating 2-arm coincidences since in the case of $\sigma_{SD}$ one arm has only a proton. For the one arm case, used for accidentals, we include $\sigma_{SD}$.

This prescription implies an energy dependence which is different than what is sometimes discussed- ie it scales with the inelastic cross section rather than being constant. Then the average number of neutrons/hemisphere per inelastic collision is 0.4 while for the proton it is ~0.5 and for $\Lambda$'s, etc. it is ~0.1. Baryon pair production is a small contribution.

The only data on 2-arm cross sections is from PHENIX. We assume that the 2-arm cross section is given by the product of the density distribution and acceptance in each arm and then test this assumption with PHENIX $\sqrt{s}$ =200, 500 GeV data.

This calculation assumes that at each beam energy the ZDC threshold is adjusted to 0.1×pbeam. Perhaps a more likely scenario is that nothing changes (gain or threshold) so the threshold is ie always 100 GeV. This doesn't change the conclusions very much.

```
MeanAccept[pbeam_] := 0.4 ∫_{0.1}^{1} Accept[pbeam x, x] ⅆx
Coinc[pbeam_] := MeanAccept[pbeam]^2
```

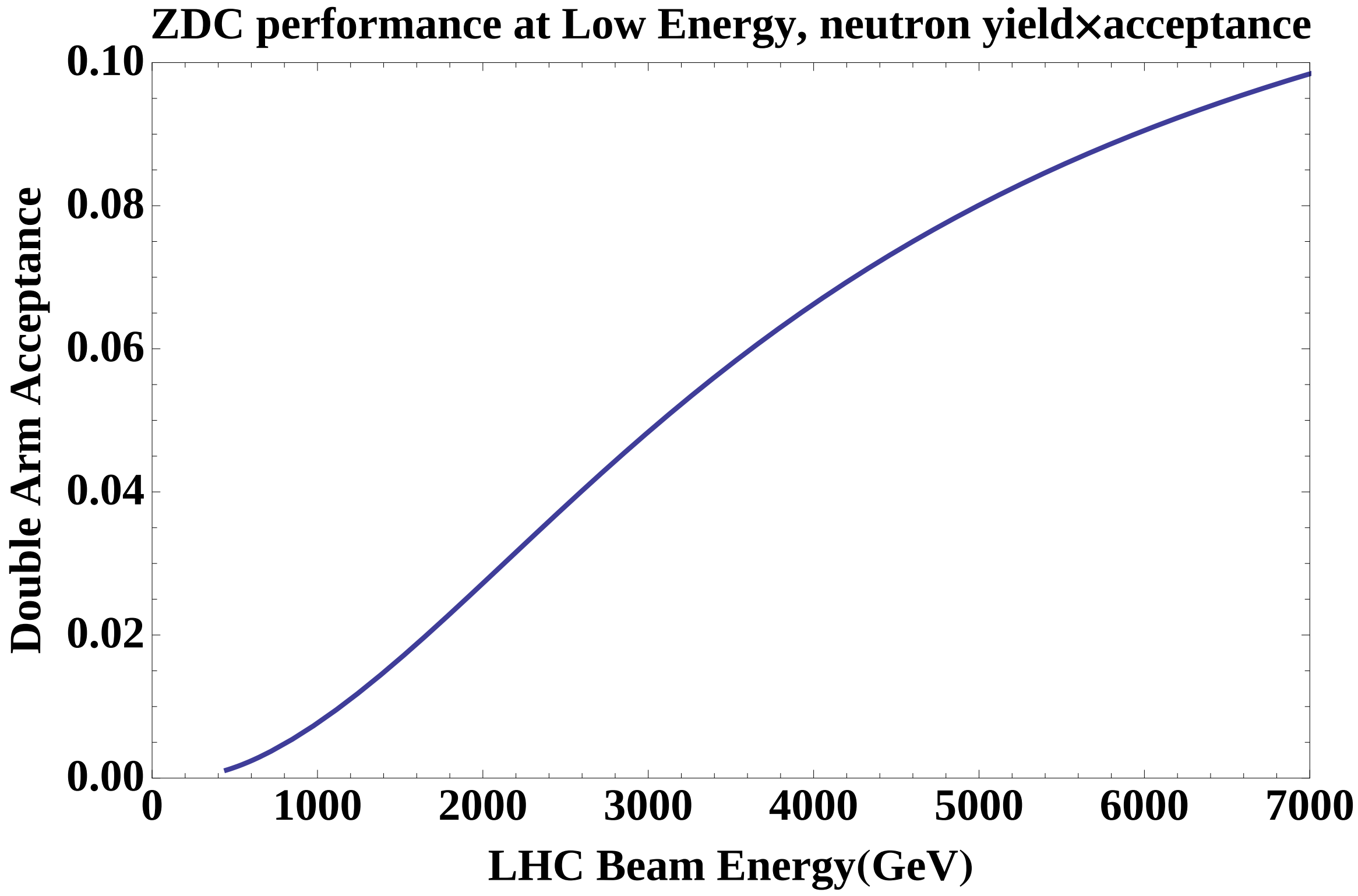

## Non - Diffractive cross sections:

$\sigma_{ND} \equiv \sigma_{total} - \sigma_{elastic} - \sigma_{SD}$ fitted to $\sqrt{p_{beam}}$

```
mb = 10^-27;
σND[pbeam_] := 37 mb + 8 * Sqrt[2 pbeam / 1000 - .9] mb
σSD = 13 mb;
Luminosity = 0.1 * 10^30;
Rate[pbeam_] := Luminosity * σND[pbeam] * Coinc[pbeam]
Rsingle[pbeam_] := Luminosity * (σND[pbeam] + σSD/2) * MeanAccept[pbeam]
```

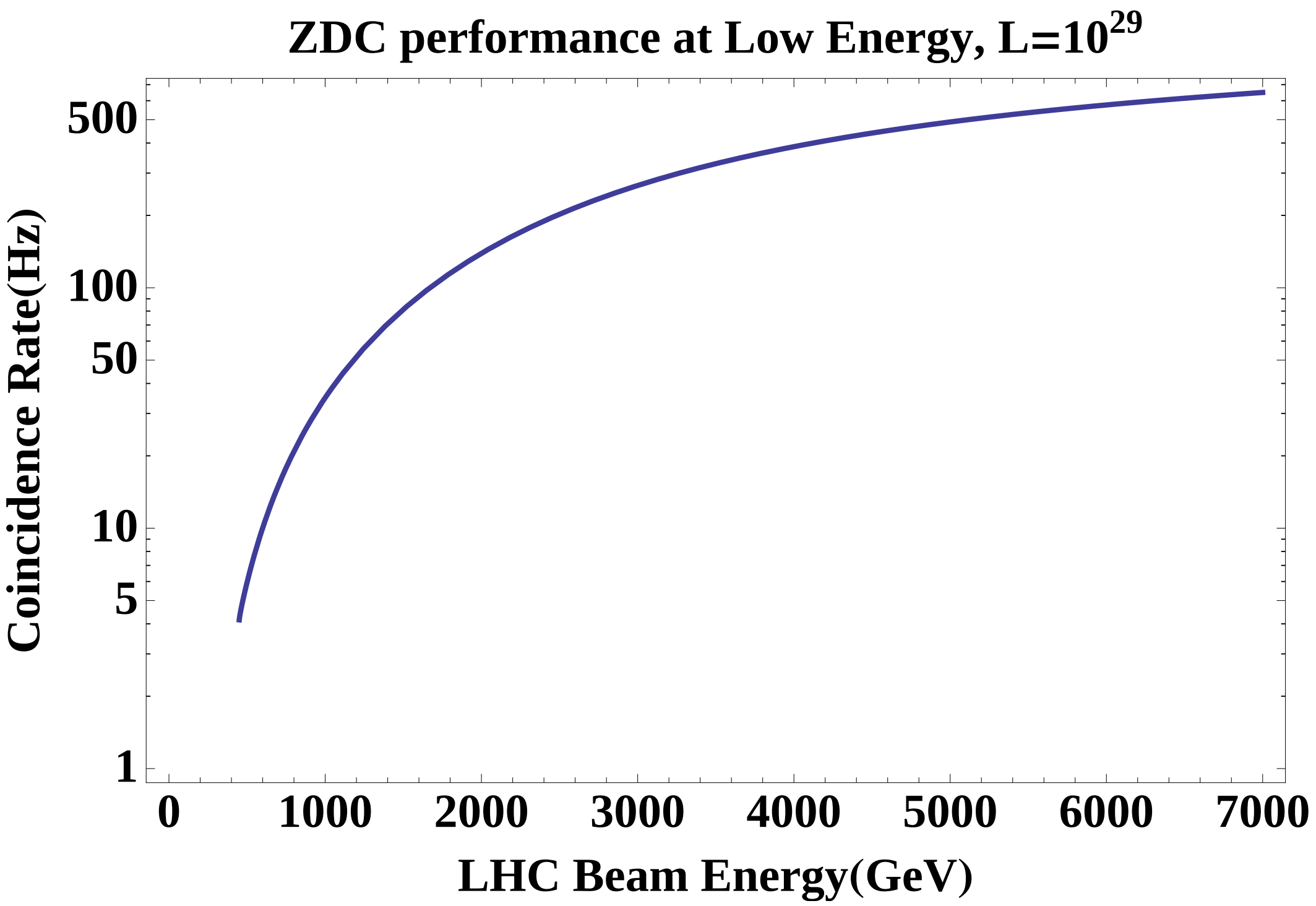

**Accidentals.**

During the first year up to 150 bunches are planned. But at the beginning the number of bunches will grow to 43 bunches. At that point the crossing geometry will change and LHCf will be removed.

The Singles rate is calculated as $\mathcal{L} \times \left(\frac{\sigma_{SD}}{2} + \sigma_{ND}\right) \times$Acceptance and the accidental rate is $2 \times$Singles $^2\tau$.

The rate of true coincidences is $\mathcal{L}\sigma_{ND}(\text{Accept})^2$.

The maximum number of bunches = 27 km/(25 nsec*30 cm/nsec) = $3 \times 10^4$/(25*.3) = 3,600. Since 3,600 corresponds to 40 Mhz, 43 bunches would give a 480 khz bunch frequency.

```
RealCoinc[Lum_] := Lum * σND[1000] * Coinc[1000];
Accid[Lum_, nbunch_] := 2 (Lum * (σND[1000] + σSD/2) * MeanAccept[1000])^2 (43/nbunch) (1/480000)
```

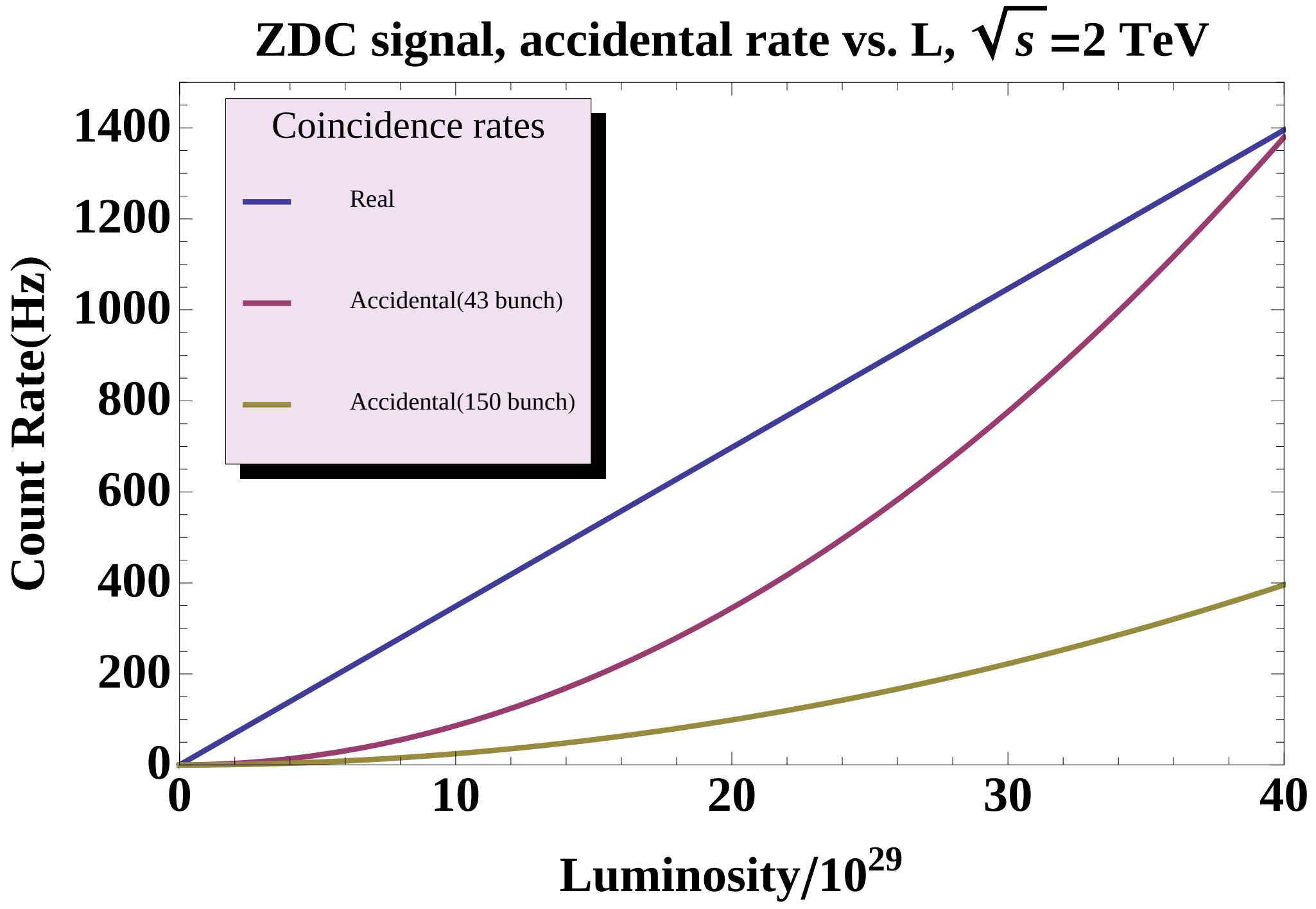

ZDC signal, accidental rate vs. L, √s =2 TeV
Coincidence rates
Real
Accidental(43 bunch)
Accidental(150 bunch)
Count Rate(Hz)
1400
1200
1000
800
600
400
200
0
0
10
20
30
40
Luminosity/10^29

**Conclusion:**

The ZDC coincidence rates will be extremely low at startup and low beam energy.

We are also likely to have trouble with accidentals already at L~2*10 $^{30}$ with 43 bunches and
at around 10 $^{31}$ with 150 bunches (this is the highest planned for the first year).

## Appendix I.Data on $p_T$ and $x_F$ distributions.

Below is the NA49 plot of $p_T$ integrated Neutron yield. Also shown are their subtractions(left).ISR distributions in $x_F$ vs $p_T$(right). Only the $p_T$ slope for $x_F$ from 0.3 to 0.7 are presented.

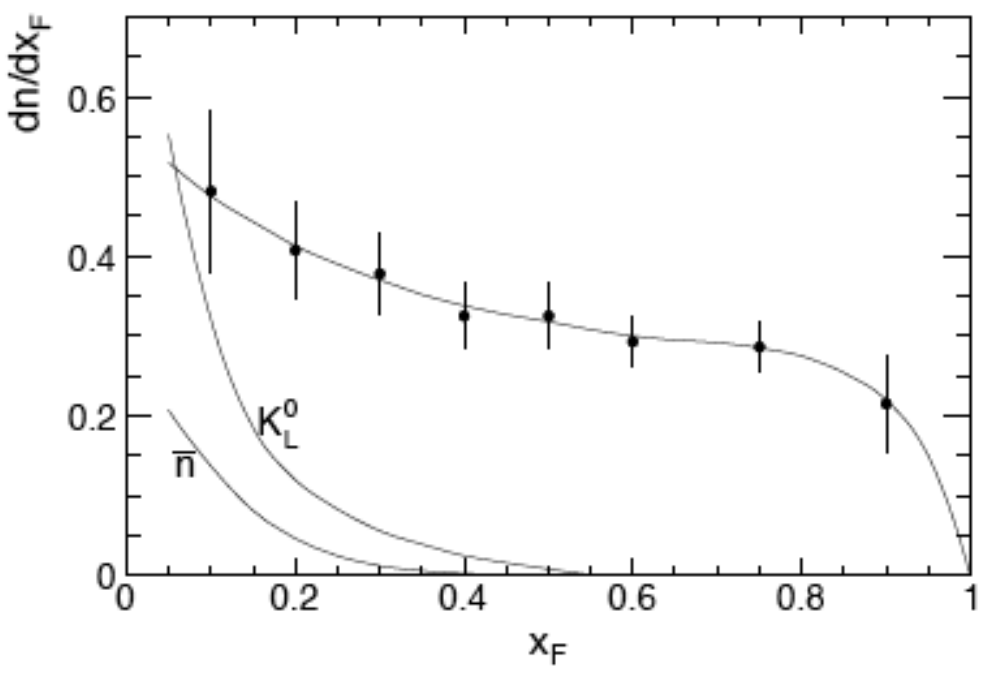

| $\langle x_L \rangle$ | $b$ GeV$^{-2}$ | $\pm(stat.)$ GeV$^{-2}$ |
|---|---|---|
| 0.49 | 4.10 | 1.10 |
| 0.54 | 4.47 | 1.00 |
| 0.58 | 3.80 | 1.10 |
| 0.62 | 6.33 | 1.00 |
| 0.65 | 8.12 | 0.90 |
| 0.69 | 8.56 | 0.90 |
| 0.73 | 6.06 | 0.80 |
| 0.76 | 9.46 | 0.90 |
| 0.80 | 8.78 | 0.80 |
| 0.84 | 12.81 | 1.10 |
| 0.88 | 9.65 | 0.90 |
| 0.95 | 8.63 | 1.10 |

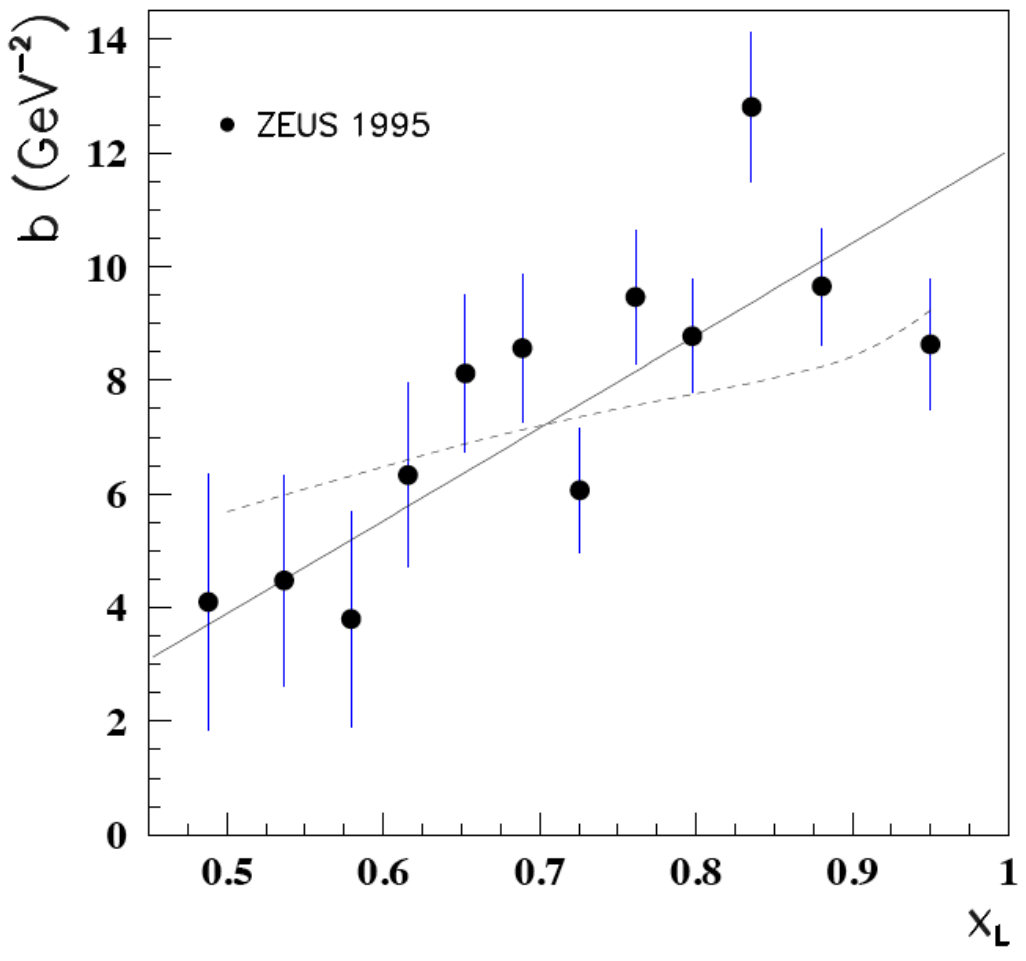

## Appendix 2: PHENIX rates and predictions

Let's check this by comparing with rates in PHENIX. At $\sqrt{s}$ =500 GeV with 859 kHz BBC rate ($\sigma$=22.9 mb) we measured a ZDC coincidence rate of 78 kHz and a singles rate of 573 kHz. We calculate 41 kHz and 488kHz. Similarly at 200 GeV we measured 3.2kHz and 64kHz and calculate 1.6 kHz and 43 kHz. The ZDC threshold was set at 15 GeV in both cases.

The agreement is pretty good considering that this model didn't start from those rates. PYTHIA simulation predicts an excess of ZDC counts due to high energy photons which could account for the small difference.

```
kHz = 10^3;
fidcutr = 5 cm / 18;
Acceptr[pl_, x_] := Final[pl * fidcutr, x]
MeanAcceptr[pbeam_] := 0.4 ∫_{0.06}^{1} Acceptr[pbeam x, x] dx
Lumi = 859 kHz / 22.9 mb;
d250 = Lumi * 30 mb * MeanAcceptr[250]^2;
s250 = 2 * Lumi * 34 mb * MeanAcceptr[250];
```

| | $\sqrt{s}$ =500 GeV coincidence | singles | $\sqrt{s}$ =200 GeV Coincidence | Singles |
|---|---|---|---|---|
| PHENIX measured | 78 000 | 573 000 | 3200. | 64 000 |
| This Calculation | 41 341.5 | 488 900. | 1614.91 | 43 112.5 |

## Appendix 3. Threshold setting.

It's likely that at low beam energy backgrounds will overwhelm the signal since it is so small. We, of course, don't know the energy spectrum of those backgrounds but it is likely, as is found in PYTHIA, that $\pi^0$ decay photons have *a* softer energy distribution than neutrons. Since the neutron acceptance rises rapidly with $x_F$, you don't lose much neutron signal if you raise the threshold during runs at injection energy. But you might suppress the photons a lot. Below we show the neutron acceptance vs threshold.

```
Thresh[y_?NumericQ] := ∫_y^1 Accept[450 x, x] dx
```

At low beam energy our conclusion is that only single arm rates can be used for luminosity monitoring. In that case $\pi^0$ photons are as good as neutrons since they are proportional also to luminosity. Then there's not much point in raising the threshold. But perhaps the machine backgrounds will also have a softer spectrum. Then a threshold of ~0.5 would be a good choice to reduce them.

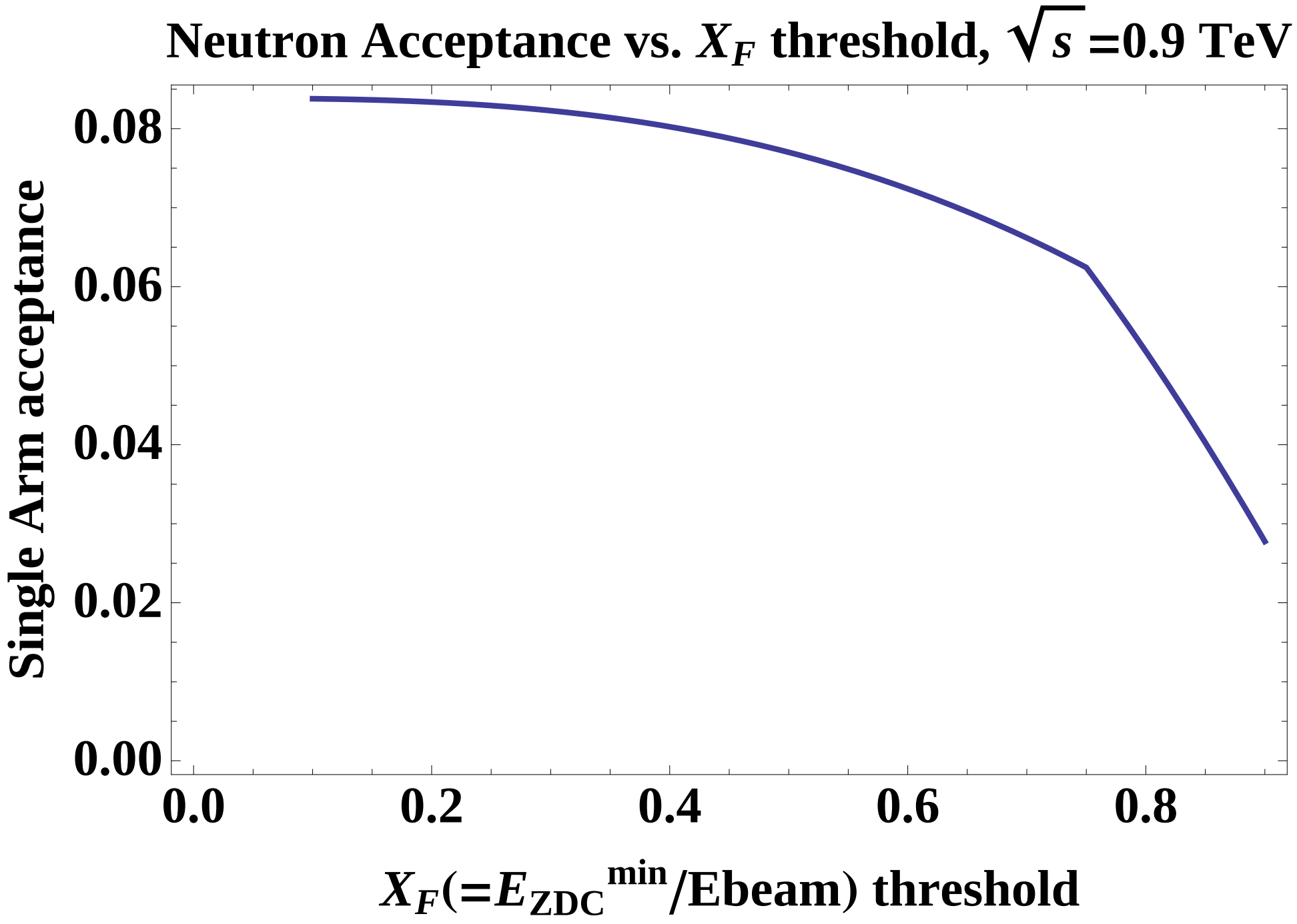

**Appendix 4. Single Arm as a luminosity Monitor:**

Since the coincidence rate is likely unviable, even at $10^{29}$, it is worth looking into single arm rates and background. Of course what is attractive about the ZDC coincidence signal at RHIC and full energy LHC is the low background.

For the single arm rate we use the same picture and expect of order 100 Hz during conditioning. There are a number of machine backgrounds which have been calculated by Mokhov but beam-gas is perhaps the most simple. We find residual gas levels of $10^{-18}\, g \big/ \text{cm}^3$ upstream of the ZDC. So the rates are:

$$\text{Rate} = \rho / m \times L \times \sigma_{\text{inel}} \times \text{freq} \times N_p \simeq 10^4 \text{ to } 10^5 \text{ Hz}$$

It should be kept in mind that not all of the NEG's will be activated during the first year so things could be worse.

Other types of background from the machine are harder to estimate but in 2009, with RHIC operating at a well known energy, a background level for the single arm of 10% was achieved when using a ZDC threshold at $x_F = 0.1$.

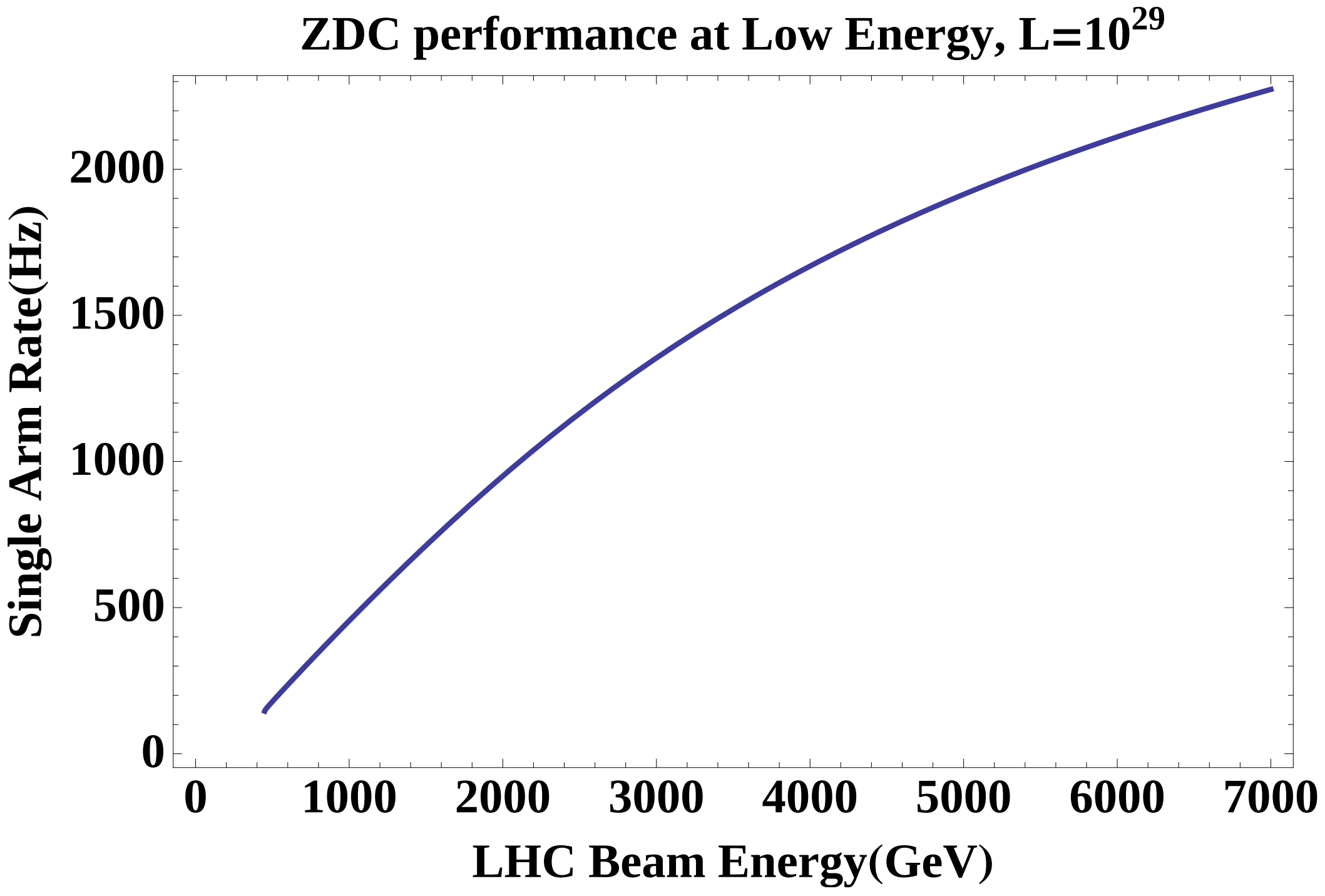